\documentclass[
    aps,
    prl,
    reprint,
    superscriptaddress,
    amsmath,
    amssymb,
    nofootinbib
]{revtex4-2}

\usepackage{graphicx}  
\usepackage{dcolumn}   
\usepackage{bm}        
\usepackage{siunitx}   
\usepackage{xcolor}    
\usepackage{hyperref}  
\usepackage{booktabs}
\usepackage{verbatim}

\usepackage{etoolbox}
\apptocmd\normalsize{%
  \abovedisplayskip=6pt plus 2pt minus 4pt
  \abovedisplayshortskip=0pt plus 2pt
  \belowdisplayskip=6pt plus 2pt minus 4pt
  \belowdisplayshortskip=4pt plus 2pt minus 2pt
}{}{}

\hypersetup{
    colorlinks=true,
    linkcolor=blue,
    citecolor=blue,
    urlcolor=blue
}

\graphicspath{{figures/}}

\begin{document}

\title{Trajectory arclength reveals thermodynamic criticality}

\author{Javier Jim\'enez-L\'opez}
\email{javier.jimenez.lopez@cern.ch}
\affiliation{
    Vrije Universiteit Brussel, Brussels, Belgium
}
\affiliation{
    Departamento de F\'isica y Matem\'aticas, Facultad de Ciencias, Universidad de Alcal\'a, 28805 Alcal\'a de Henares, Madrid, Spain.
}

\author{V. J. Garc\'ia-Garrido}
\email{vjose.garcia@uah.es (corresponding author)}
\affiliation{
    Departamento de F\'isica y Matem\'aticas, Facultad de Ciencias, Universidad de Alcal\'a, 28805 Alcal\'a de Henares, Madrid, Spain.
}

\date{\today}


\begin{abstract}

Thermodynamic criticality arises at continuous phase transitions, where collective interactions produce non-analytic behavior in equilibrium thermodynamic quantities in the thermodynamic limit. We show that this criticality also leaves an analytically predictable signature in the arclength of microscopic Hamiltonian trajectories. For the attractive Hamiltonian mean-field model, we derive the thermodynamic limit of the equilibrium-averaged arclength, formulated as a normalized Lagrangian descriptor, and demonstrate that it develops a kink precisely at the phase transition. We trace this non-analyticity to spontaneous magnetization and derive the leading critical finite-size correction, scaling as $N^{-1/2}$ with an explicit analytical prefactor. Simulations with up to $10^5$ degrees of freedom confirm both predictions without adjustable parameters. Requiring neither tangent-space evolution nor prior specification of an order parameter for its computation, trajectory arclength provides a simple, quantitative link between microscopic Hamiltonian motion and collective thermodynamic criticality.

\end{abstract}

\maketitle

\textit{Introduction.---}Critical phenomena reveal how interactions among microscopic degrees of freedom give rise to non-analytic macroscopic behavior in the thermodynamic limit \cite{Huang1987,Wilson1971}. However, numerical simulations necessarily involve finite systems and finite observation times, motivating the search for microscopic observables that retain identifiable signatures of criticality. A fundamental question is whether such signatures can be extracted directly from Hamiltonian trajectories and characterized analytically.

The connection between microscopic Hamiltonian dynamics and thermodynamic phase transitions has been extensively investigated through chaotic instability. Caiani \emph{et al.} identified characteristic behavior of Lyapunov exponents and geometric observables near the transition in three-dimensional classical $XY$ models \cite{Caiani1997}, with related findings in lattice $\phi^4$ models and long-range Hamiltonian systems \cite{Caiani1998PRE,Latora1998,Latora1999}. The geometric approach developed by Pettini and collaborators relates Hamiltonian instability to curvature fluctuations of the underlying mechanical manifolds \cite{Casetti2000,Pettini2024}. For the Hamiltonian mean-field (HMF) model, analytical estimates and non-trivial finite-size scaling laws have been obtained for the largest Lyapunov exponent \cite{Firpo1998,Ginelli2011}, while complementary studies have connected its equilibrium transition to changes in configuration-space topology \cite{Casetti1999}. Independently, microcanonical approaches have established how thermodynamic quantities can be recovered from dynamical averages involving the Hamiltonian and its derivatives \cite{Rugh1997,Franzosi2012}. These results demonstrate the close relationship between Hamiltonian dynamics and equilibrium thermodynamics, but conventional Lyapunov-exponent calculations require the evolution of tangent-space dynamics.

A more direct possibility is to examine the arclength accumulated by Hamiltonian trajectories. Lagrangian descriptors (LDs), constructed by integrating scalar observables along trajectories, have been used to uncover invariant structures in phase space \cite{Mancho2013,WigginsGarciaGarrido2022}. More recently, normalized time-averaged phase space arclength was shown to distinguish regular and chaotic motion through its finite-time convergence and fluctuation properties under suitable assumptions \cite{JimenezLopez2026}. For a Hamiltonian flow $\dot{\mathbf{z}}=X_H(\mathbf{z}) =\mathbb{J}\nabla H(\mathbf{z})$, the Euclidean speed in fixed dimensionless canonical coordinates satisfies $\|X_H\|=\|\nabla H\|$, providing a direct connection between a trajectory-based observable and equilibrium averages over constant-energy surfaces. This raises a natural question: can such a simple observable also encode collective thermodynamic criticality?

This work provides a positive answer to this question by establishing an analytically controlled connection between microscopic Hamiltonian dynamics and equilibrium criticality in the attractive HMF model \cite{AntoniRuffo1995}. Whereas previous approaches have associated phase transitions with Lyapunov instability and changes in configuration-space topology, we show that criticality is encoded in a directly computable trajectory observable: time-averaged phase-space arclength. By expressing the instantaneous squared speed exactly in terms of the conserved energy and collective observables, we derive its equilibrium thermodynamic limit and demonstrate that it develops a kink precisely at the critical energy. Remarkably, this non-analyticity originates in the onset of spontaneous magnetization, despite the smoothness of the underlying Hamiltonian flow. We further derive the leading critical finite-size correction, which scales as $N^{-1/2}$ with an explicit analytical prefactor and no adjustable parameters. These results establish a direct quantitative link between microscopic trajectory motion, equilibrium symmetry breaking, and critical fluctuations. Since the evaluation of this observable requires neither tangent-space propagation nor prior knowledge of an order parameter, our findings provide a computationally simple route to probing equilibrium criticality in many-body Hamiltonian systems.

\textit{Hamiltonian arclength descriptor.---}Lagrangian descriptors are trajectory-integrated observables whose dependence on initial conditions reveals invariant structures in phase space \cite{Mancho2013,Lopesino2017}. Their spatial variations have also been exploited to distinguish regular and chaotic dynamics through comparisons between neighboring trajectories \cite{Hillebrand2022,Zimper2023,JimenezLopezDelta2026} or derivatives with respect to initial conditions \cite{Daquin2022,Caliman2025}.

Consider an autonomous Hamiltonian $H_N(\mathbf{q},\mathbf{p})$ with $N$ degrees of freedom. In fixed dimensionless canonical coordinates $\mathbf{x}=(\mathbf{q},\mathbf{p})$, Hamilton's equations take the form:
\begin{equation}
\dot{\mathbf{x}}
=X_{H_N}(\mathbf{x})
=J \, \nabla H_N(\mathbf{x}),
\qquad
J=
\begin{pmatrix}
0&I_N\\
-I_N&0
\end{pmatrix},
\end{equation}
where $X_{H_N}$ is the Hamiltonian vector field and $I_N$ is the identity matrix of order $N$. We consider the forward arclength LD and its normalized time average:
\begin{equation}
\begin{aligned}
\mathcal{L}_f(\mathbf{x}_0,\tau)
&=\int_0^\tau
\left\|X_{H_N}(\mathbf{x}(t;\mathbf{x}_0))\right\|
\,\mathrm{d}t,\\
\Omega_N(\mathbf{x}_0,\tau)
&=\frac{\mathcal{L}_f(\mathbf{x}_0,\tau)}
{\tau\sqrt{N}},
\end{aligned}
\label{eq:LD_Omega_definition}
\end{equation}
where $\|\cdot\|$ denotes the Euclidean norm in the chosen coordinates. The factor $N^{-1/2}$ yields an intensive observable when the squared phase space speed is extensive.

Previous work showed that the finite-time convergence and fluctuation properties of this observable provide information about regular and chaotic dynamics \cite{JimenezLopez2026}. Here we investigate a different aspect: the dependence of its equilibrium expectation on energy and system size. For the invariant microcanonical measure $\mu_{N,e}$ at energy per particle $e=H_N/N$, stationarity yields the exact identity:
\begin{equation}
\left\langle\Omega_N(\cdot,\tau)\right\rangle_{\mu_{N,e}}
=\frac{1}{\sqrt{N}}
\left\langle|X_{H_N}|\right\rangle_{\mu_{N,e}} \,,
\label{eq:Omega_equilibrium}
\end{equation}
valid for any $\tau>0$, independently of ergodicity.

Computationally, evaluating $\Omega_N$ requires only the integration of Hamilton's equations and a scalar accumulator. Once the Hamiltonian vector field is available, computing its Euclidean norm adds $O(N)$ operations per sampled state, without the need to propagate tangent-space dynamics. Unlike Lyapunov-exponent calculations, which require integrating variational equations \cite{Skokos2010a}, the arclength descriptor is straightforward to implement for systems with many degrees of freedom and large
trajectory ensembles. Its overall computational cost, however, depends on the Hamiltonian and the integration time required for statistical convergence.

\textit{Hamiltonian mean-field model.---}The Hamiltonian mean-field model is a paradigmatic system for investigating collective phenomena, Hamiltonian chaos, and equilibrium phase transitions in systems with long-range interactions \cite{AntoniRuffo1995,Chavanis2005,CAMPA200957}. It consists of $N$ globally coupled classical rotators, with angular coordinates $\theta_i\in[0,2\pi)$ and conjugate momenta $p_i$, governed by the Hamiltonian:
\begin{equation}
H_N(\boldsymbol{\theta},\mathbf{p})
=\sum_{i=1}^{N}\frac{p_i^2}{2}
+\frac{J}{2N}\sum_{i,j=1}^{N}
\left[1-\cos(\theta_i-\theta_j)\right],
\label{eq:HMF_og}
\end{equation}
where $J$ determines the interaction strength and sign, and the $1/N$ normalization ensures an extensive energy. For attractive interactions ($J>0$), the model undergoes a continuous equilibrium phase transition from a low-energy, magnetized phase to a high-energy, homogeneous phase \cite{AntoniRuffo1995,Dauxois2002,Dauxois2000}. In the zero-total-momentum frame, the critical temperature is $T_c=J/2$, corresponding to the critical energy density $\mathcal{U}_c=3J/4$. The relationship between this transition and microscopic chaos has been investigated through Lyapunov exponents and geometric properties of the Hamiltonian flow \cite{Latora1998,Firpo1998,Casetti1999}. In the collisionless mean-field limit, the dynamics is governed by the Vlasov equation, which also provides the framework for understanding long-lived
quasi-stationary states
\cite{Antoniazzi2007,Barre2006Vlasov,Braun1977,Spohn1991}.

Here we restrict our attention to the attractive case and establish an analytical connection between equilibrium criticality and phase space trajectory arclength. Our approach expresses the instantaneous squared speed of the Hamiltonian flow in terms of collective observables, then exploits equilibrium concentration to determine its thermodynamic limit and a large-deviation expansion to derive the leading critical finite-size correction \cite{Barre2005,Touchette2009}.

\textit{Theoretical derivation.---}Let us first introduce the magnetization vector:
\begin{equation}
\mathbf{M}
=\frac{1}{N}\sum_{i=1}^{N} (\cos\theta_i,\sin\theta_i)
=M(\cos\phi,\sin\phi) \,,
\end{equation}
where $M$ and $\phi$ denote its magnitude and orientation, respectively. Using the identity:
\begin{equation}
    \sum_{i,j}\cos(\theta_i-\theta_j)=N^2M^2 \;,
\end{equation}
we can rewrite the Hamiltonian as:
\begin{equation}
H_N=\sum_{i=1}^{N}\frac{p_i^2}{2}
+\frac{JN}{2}(1-M^2) \,.
\label{eq:HMF_v2}
\end{equation}
The energy density $\mathcal{U}=H_N/N$ is conserved, whereas the magnetization generally evolves along Hamiltonian trajectories. The equations of motion read:
\begin{equation}
\dot{\theta}_i=p_i,
\quad
\dot{p}_i=-JM\sin(\theta_i-\phi) \,,
\label{eq:motion}
\end{equation}
and the normalized arclength descriptor becomes:
\begin{equation}
\Omega_N(\tau)=\dfrac{1}{\tau}\int_0^\tau
\left[
\dfrac{1}{N}\sum_{i=1}^{N}
\left(\dot{\theta}_i^2+\dot{p}_i^2\right)
\right]^{1/2}\mathrm{d}t \,.
\label{eq:Omega_HMF}
\end{equation}
From Hamilton's equations and energy conservation, the instantaneous squared phase space speed satisfies the exact identity:
\begin{equation}
\frac{\|X_{H_N}\|^2}{N}
=2\mathcal{U}-J
+\left(J+J^2 A_N\right)M^2,
\label{eq:HMF_speed_identity}
\end{equation}
where:
\begin{equation}
A_N=\frac{1}{N}\sum_{i=1}^{N}
\sin^2(\theta_i-\phi).
\end{equation}
Hence, the arclength descriptor can be expressed entirely in terms of the conserved energy density and collective observables:
\begin{equation}
\Omega_N(\tau)=\frac{1}{\tau}\int_0^\tau
\sqrt{2\mathcal{U}-J+
\left[J+J^2A_N\right]M^2}
\,\mathrm{d}t.
\label{eq:Omega_HMF_collective}
\end{equation}
This exact relation provides the starting point for connecting the equilibrium behavior of the magnetization to the thermodynamic limit and critical finite-size corrections of $\Omega_N$.

In the thermodynamic limit $N \to \infty$ the mean-field description becomes exact \cite{Messer1982, BraunHepp1977} and the equilibrium state, which is the same in the canonical and microcanonical ensembles, is described by the single-particle distribution \cite{AntoniRuffo1995, Inagaki1993, Chavanis2005, chavanis2010, Barre2005, CAMPA200957, Manos02122011}:
\begin{equation}
\begin{aligned}
    f_{eq} (\theta, p) = & \dfrac{\sqrt{\beta}}{(2\pi)^{3/2} I_0(\beta J M)} \\
    & \times\exp \left[ - \beta \left( \dfrac{p^2}{2} - J M \cos(\theta - \phi)  \right) \right]
\end{aligned}\, ,
\end{equation}
where $\beta = T^{-1}$ is the inverse temperature and $I_n$ is the modified Bessel function of order $n$. We assume throughout that the system has relaxed to this state, i.e., that it is not trapped in a quasi-stationary state \cite{Yamaguchi2004}. The distribution factorises into a Gaussian for the momenta, with $\langle p^2 \rangle = 1/\beta$, and:
\begin{equation}\label{eq:ang_dist}
    \rho (\theta) = \dfrac{\exp(\beta J M \cos(\theta - \phi))}{2\pi I_0(\beta J M)} \, ,
\end{equation}
for the angles, which is peaked at $\phi$ for $M > 0$ and uniform for $M = 0$. Self-consistency requires:
\begin{equation}\label{eq:consistency}
    M = \langle \cos(\theta - \phi) \rangle = \dfrac{I_1(z)}{I_0(z)} \, ,
\end{equation}
with $z = \beta J M$. On the ordered branch ($M>0$), using $I_0(z) - I_2(z) = (2/z) I_1(z)$ \cite{NIST:DLMF}:
\begin{equation}
    A_N \to \langle \sin^2 (\theta - \phi) \rangle = \dfrac{1}{2} \left[ 1 - \dfrac{I_2(z)}{I_0(z)} \right] = \dfrac{1}{\beta J} \, .
\end{equation}

On the homogeneous branch $A_N\to1/2$ and $M\to0$. In both phases, the equilibrium concentration yields:
\begin{equation}
\begin{split}
        \Omega_{\infty} &= \sqrt{\langle p^2 \rangle + J^2 M^2 \langle \sin^2 (\theta - \phi) \rangle} \\[.1cm] &= \sqrt{T (1 + J M^2)} \, ,
\end{split}
\end{equation}
and eliminating the temperature through $\mathcal{U} = T/2 + (J/2) (1 - M^2)$ we obtain:
\begin{equation} \label{eq:Omega_MU}
    \Omega_{\infty} = \sqrt{(2\mathcal{U} - J + J M^2)(1 + J M^2)} \, .
\end{equation}

The critical point follows from Eq.\eqref{eq:consistency}. Since $I_1(z)/I_0(z) = z/2 - z^3/16 + \mathcal{O}(z^5)$, a solution with $M > 0$ exists only if $\beta J > 2$, and the system undergoes a second order phase transition at \cite{AntoniRuffo1995, CAMPA200957}:
\begin{equation}
    T_c = \dfrac{J}{2} \, , \hspace{0.5cm} \mathcal{U}_c = \dfrac{3J}{4} \, ,
\end{equation}
with $M > 0$ for $\mathcal{U} < \mathcal{U}_c$ and $M = 0$ for $\mathcal{U} \geq \mathcal{U}_c$. Hence, $\Omega_{\infty}$ is given by Eq.\eqref{eq:Omega_MU} below $\mathcal{U}_c$ and by:
\begin{equation} \label{eq:Omega_U}
    \Omega_{\infty}(\mathcal{U}) = \sqrt{2\mathcal{U} - J} \hspace{0.5cm} \mathrm{if} \hspace{0.25cm} \mathcal{U} \geq \mathcal{U}_c \, ,
\end{equation}
which is continuous at the transition, where $\Omega_{\infty}(\mathcal{U}_c) = \sqrt{J / 2}$. However, its derivative is discontinuous. Above $\mathcal{U}_c$, Eq.\eqref{eq:Omega_U} gives $\mathrm{d}\Omega_{\infty} / \mathrm{d}\mathcal{U} = (2\mathcal{U} - J)^{-1/2}$. Below it, inverting the expansion of Eq.\eqref{eq:consistency} gives $z = 2M + M^3 + \mathcal{O}(M^5)$, i.e., $\beta J = 2 + M^2 + \mathcal{O}(M^4)$, and therefore:
\begin{equation}
    \mathcal{U} = \mathcal{U}_c - \dfrac{5J}{8} M^2 + \mathcal{O}(M^4) \, ,
\end{equation}
so that, defining $\delta = \mathcal{U}_c - \mathcal{U} > 0$:
\begin{equation}
    \Omega^2_{\infty} = T (1 + J M^2)  = \dfrac{J}{2} + \dfrac{2}{5} (2J - 1) \, \delta + \mathcal{O}(\delta^2) \, .
\end{equation}

The one-sided derivatives at the critical energy are:
\begin{equation}
    \left. \dfrac{\mathrm{d}\Omega_{\infty}}{\mathrm{d}\mathcal{U}} \right|_{\mathcal{U}^{-}_c} = \dfrac{\sqrt{2} \, (1 - 2J)}{5 \sqrt{J}} \, , \hspace{0.5cm}
    \left. \dfrac{\mathrm{d}\Omega_{\infty}}{\mathrm{d}\mathcal{U}} \right|_{\mathcal{U}^{+}_c} = \sqrt{\dfrac{2}{J}} \, ,
\end{equation}
and their difference:
\begin{equation}
    \left. \dfrac{\mathrm{d}\Omega_{\infty}}{\mathrm{d}\mathcal{U}} \right|_{\mathcal{U}^{+}_c} - \left. \dfrac{\mathrm{d}\Omega_{\infty}}{\mathrm{d}\mathcal{U}} \right|_{\mathcal{U}^{-}_c} = \sqrt{\dfrac{2}{J}} \, \dfrac{4 + 2J}{5} \, ,
\end{equation}
is strictly positive for every $J > 0$: $\Omega_{\infty}$ has a kink at $\mathcal{U}_c$ and therefore reveals the phase transition.

Since the thermodynamic limit cannot be reached numerically, we now derive how $\Omega_N$ approaches $\Omega_{\infty}$ at the critical energy. For $\tau \to \infty$, the time average in Eq.\eqref{eq:Omega_HMF} is replaced by the microcanonical average:
\begin{equation}
    \langle \Omega_N \rangle = \left\langle \sqrt{2\mathcal{U} - J + J M_N^2 (1 + J A_N)} \right\rangle \, .
\end{equation}
At $\mathcal{U}_c$ the magnetization scales as $M_N \sim N^{-1/4}$ \cite{Botet1982, Botet1983, Ginelli2011}, which also follows from Eq.\eqref{eq:P_M} below. Expanding the square root in powers of $M_N^2$, with $2\mathcal{U}_c - J = J/2$ and $A_N \to 1/2$, we obtain:
\begin{equation} \label{eq:Omega_N_M2}
    \begin{aligned}
         \langle \Omega_{N}(\mathcal{U}_c) \rangle = \sqrt{\dfrac{J}{2}} + \sqrt{\dfrac{J}{2}} \left( 1 + \dfrac{J}{2} \right) \langle M^2_N(\mathcal{U}_c) \rangle & \\+ o(N^{-1/2}) \, ,
    \end{aligned}
\end{equation}
where the neglected terms are of order $\langle M^4_N \rangle \sim N^{-1}$ and $\langle M^2_N (A_N - 1/2) \rangle$. To obtain $\langle M^2_N(\mathcal{U}_c)\rangle$, we use the microcanonical magnetization distribution at fixed $P=\sum_i p_i=0$, as in the simulations. At fixed $\mathcal{U}$ and $M$, $K=(N/2)(2\mathcal{U}-J+JM^2)$. The momentum constraint removes one independent component, giving:
\begin{equation}
    \int \delta\!\left(\sum_i p_i\right)\delta\!\left(K-\frac12\sum_i p_i^2\right) \; \mathrm{d}^N p  \propto K^{(N-3)/2}\, ,
\end{equation}
which gives, to leading order in $N$, the kinetic entropy per particle:
\begin{equation}
    s_K (\mathcal{U}, M)= \dfrac{1}{2} \log (2\mathcal{U} - J + J M^2) + \alpha \, ,
\end{equation}
where $\alpha$ is independent of $M$ and $\mathcal{U}$. The momentum constraint changes subleading terms but not the leading critical amplitude below.

The number of angular configurations with magnetization $M$ is determined by the most probable angular distribution at fixed $M$, which has the form of Eq.\eqref{eq:ang_dist} with $\beta J M$ replaced by a Lagrange multiplier $z$ fixed by $M = I_1(z) / I_0(z)$. The configurational entropy per particle is thus:
\begin{equation}
    \begin{aligned}
            s_{\mathrm{conf}}(M) &= - \int_{0}^{2\pi}   \rho(\theta) \log \rho(\theta) \; \mathrm{d}\theta \\  &= \log \left[2\pi I_0(z)\right] - z M \, .
    \end{aligned}
\end{equation}
Using $z = 2M + M^3 + \mathcal{O}(M^5)$ we get:
\begin{equation}
    s_{\mathrm{conf}}(M) = \log(2\pi) - M^2 - \dfrac{1}{4}M^4 + \mathcal{O}(M^6) \, .
\end{equation}

The total entropy per particle is $s = s_K + s_{\mathrm{conf}}$. At the critical energy $s_K = \frac{1}{2} \log (J/2) + M^2 - M^4 + \alpha + \mathcal{O}(M^6)$, so the terms in $M^2$ cancel exactly and:
\begin{equation}
    s(\mathcal{U}_c, M) = s(\mathcal{U}_c, 0) - \dfrac{5}{4}M^4 + \mathcal{O}(M^6) \, ,
\end{equation}
independently of $J$. Since to exponentially leading order $P_N(\mathbf{M}) \propto \exp \left[ N s(\mathcal{U}, M) \right]$ \cite{Touchette2009}, at $\mathcal{U} = \mathcal{U}_c$:
\begin{equation} \label{eq:P_M}
    P_N(\mathbf{M}) \propto \exp \left( - \dfrac{5N}{4} M^4 \right) \, ,
\end{equation}
at leading order in the critical window $M=O(N^{-1/4})$; higher entropy terms and smooth prefactors give subleading corrections. Since $\mathrm{d}^2\mathbf{M}=M\,\mathrm{d}M\,\mathrm{d}\phi$,
\begin{equation}
    \langle  M^2_N (\mathcal{U}_c) \rangle \sim \dfrac{\int_0^{\infty} \mathrm{d}M \, M^3 \, e^{-5NM^4/4}}{\int_0^{\infty} \mathrm{d}M \, M \, e^{-5NM^4/4}} = \dfrac{2}{\sqrt{5 \pi N}} \, ,
\end{equation}
which, substituted in Eq.\eqref{eq:Omega_N_M2}, gives the final result:
\begin{equation} \label{eq:asym_pred}
    \langle \Omega_{N}(\mathcal{U}_c) \rangle = \sqrt{\dfrac{J}{2}} + \dfrac{(2 + J) \sqrt{J}}{\sqrt{10 \pi N}} + o(N^{-1/2}) \, .
\end{equation}

\textit{Results.---}All results are computed in dimensionless phase space
coordinates, with unit rotor inertia and $J=1$. The LD is defined using the Euclidean norm in these coordinates. Initial conditions are generated by independent microcanonical Monte Carlo chains at fixed $H=N\mathcal{U}$ and zero total momentum $P=\sum_i p_i=0$. Each trajectory is then integrated for $\tau=2\cdot10^4$ time units with step $h=0.05$ using a symplectic scheme~\cite{Yoshida1990}. The brackets in the
figures denote averages of $\Omega_N(\tau)$ over independent replicas, and error bars represent $95\%$ confidence intervals.

\begin{figure}[htpb]
    \centering
    \includegraphics[scale=0.91]{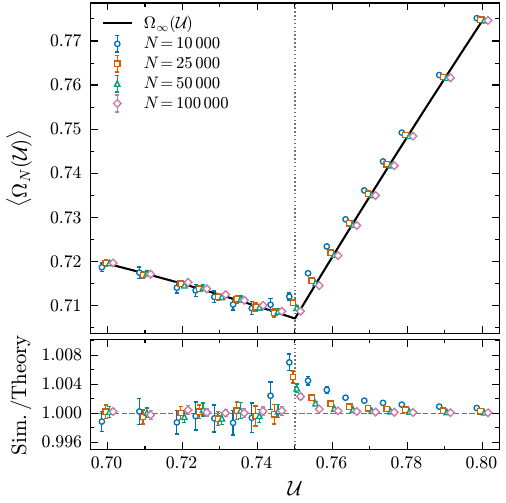}
    \caption{Energy dependence of $\langle\Omega_N(\mathcal{U})\rangle$ for $J=1$ and different system sizes $N$ near the critical energy $\mathcal{U}_c=0.75$, marked by the vertical dotted lines. The lower panel shows the ratio of the simulation results to the theoretical prediction given by Eqs.\eqref{eq:Omega_MU} and \eqref{eq:Omega_U}. Each point averages $64$ independent replicas, each integrated for $\tau=2\cdot10^4$ time units. Error bars indicate $95\%$ confidence intervals. Symbols are displaced horizontally for visual clarity.}
    \label{fig:U_dep}
\end{figure}

Figure~\ref{fig:U_dep} shows the energy dependence of
$\langle\Omega_N\rangle$ together with the thermodynamic
prediction for several system sizes. The simulations reproduce the distinct behavior on the two sides of the transition: $\langle\Omega_N\rangle$ decreases throughout the ordered phase and increases in the homogeneous phase, converging toward the two analytical branches given by Eqs.~\eqref{eq:Omega_MU} and \eqref{eq:Omega_U}. Their intersection at $\mathcal{U}_c$ produces the kink of the thermodynamic-limit curve. At finite $N$, this non-analyticity is necessarily rounded by critical fluctuations, yielding instead a smooth minimum in the vicinity of $\mathcal{U}_c$. The excess of the numerical averages above the thermodynamic prediction, shown in the
lower panel, is largest in this critical region and systematically decreases with increasing $N$. Thus the finite-size data display the expected crossover from a smooth finite-$N$ observable to the non-differentiable thermodynamic limit. Importantly, the finite-$N$ minimum
is not itself the phase-transition criterion; rather, its progressive sharpening with increasing system size approaches the analytically predicted non-analyticity at $\mathcal{U}_c$.

\begin{figure}[htpb]
    \centering
    \includegraphics[scale=0.91]{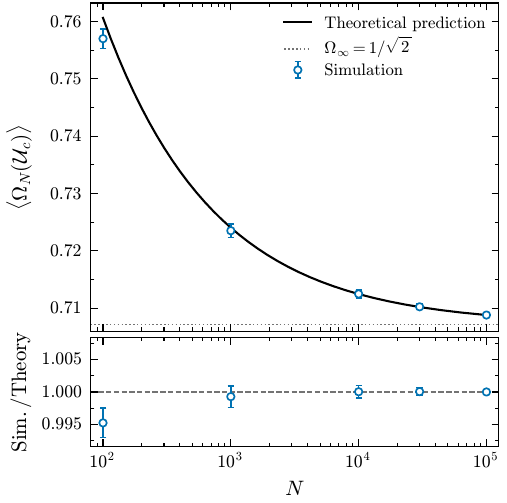}
    \caption{System-size dependence of $\langle\Omega_N(\mathcal{U}_c)\rangle$ for $J=1$ at the critical energy $\mathcal{U}_c=0.75$. The solid line shows the leading asymptotic prediction given by Eq.\eqref{eq:asym_pred}, while the horizontal dotted line marks the thermodynamic limit $1/\sqrt{2}$. The lower panel shows the ratio of the simulation results to the asymptotic prediction. Each point averages $100$ independent replicas, each integrated for $\tau=2\cdot10^4$ time units. Error bars indicate $95\%$ confidence intervals.}
    \label{fig:N_dep}
\end{figure}

At the critical energy, this rounding can be tested quantitatively. Figure~\ref{fig:N_dep} compares $\langle\Omega_N(\mathcal{U}_c)\rangle$ with the asymptotic prediction of Eq.~\eqref{eq:asym_pred}. The averages
approach the thermodynamic value $1/\sqrt{2}$ from below, with a leading correction proportional to $N^{-1/2}$. For $N\geq10^3$, the numerical results agree with the analytical expression within their confidence intervals, whereas the small deviation observed for $N=100$ is
consistent with subleading finite-size contributions. Importantly, neither the exponent nor the amplitude is fitted to the simulations: both follow from the microcanonical critical fluctuations derived above. The agreement therefore provides a direct test of the mechanism linking critical magnetization fluctuations to the geometry of Hamiltonian trajectories, rather than only of an empirical finite-size scaling law.

\textit{Conclusions.---}We have established an analytical connection between microscopic Hamiltonian motion and thermodynamic criticality through time-averaged phase space trajectory arclength. For the attractive Hamiltonian mean-field model, we show that the thermodynamic limit of this observable develops a kink precisely at the equilibrium phase transition. Crucially, this non-analyticity originates in the onset of spontaneous magnetization, rather than any singularity in the underlying Hamiltonian equations of motion. An exact relation between phase space speed and collective observables reveals this mechanism and connects critical magnetization fluctuations to the leading finite-size correction, which we derive analytically as an $N^{-1/2}$ law with an explicit prefactor. Numerical simulations involving up to $10^5$ particles confirm both predictions without adjustable parameters.

More broadly, our results demonstrate how equilibrium ordering and critical fluctuations become encoded in a simple single-trajectory observable. Unlike Lyapunov-based approaches, the arclength descriptor requires neither tangent-space propagation nor prior identification of an order parameter for its computation. Although the manifestation of critical singularities must be established for each system, the underlying analytical framework offers a route to investigating phase transitions in more general many-body Hamiltonian systems. Together with its established applications to dynamical characterization, these findings position trajectory arclength as a promising link between microscopic dynamics and emergent collective behavior.

\textit{Acknowledgments.---}The authors thankfully acknowledge Prof. Sáez Landete of the Universidad de Alcalá for providing the computational resources necessary to carry out the numerical simulations supporting the results presented in this work.


\bibliography{references}

@article{Casetti1999,
  author  = {Casetti, Lapo and Cohen, E. G. D. and Pettini, Marco},
  title   = {Topological Origin of the Phase Transition in a Mean-Field Model},
  journal = {Physical Review Letters},
  volume  = {82},
  number  = {21},
  pages   = {4160--4163},
  year    = {1999},
  doi     = {10.1103/PhysRevLett.82.4160},
  publisher = {American Physical Society}
}

@article{Latora1998,
  author = {Latora, Vito and Rapisarda, Andrea and Ruffo, Stefano},
  title = {Lyapunov Instability and Finite Size Effects in a System with Long-Range Forces},
  journal = {Physical Review Letters},
  volume = {80},
  pages = {692--695},
  year = {1998},
  doi = {10.1103/PhysRevLett.80.692}
}

@article{Latora1999,
  author = {Latora, Vito and Rapisarda, Andrea and Ruffo, Stefano},
  title = {Chaos and Statistical Mechanics in the {H}amiltonian Mean Field Model},
  journal = {Physica D: Nonlinear Phenomena},
  volume = {131},
  pages = {38--54},
  year = {1999},
  doi = {10.1016/S0167-2789(98)00217-6}
}

@Inbook{Skokos2010a,
author="Skokos, Ch.",
editor="Souchay, Jean J.
and Dvorak, Rudolf",
title={{The {L}yapunov {C}haracteristic {E}xponents and their computation}},
bookTitle="Dynamics of Small Solar System Bodies and Exoplanets",
year="2010",
publisher="Lectures Notes in Physics, vol 790. Springer",
address="Berlin, Heidelberg",
pages="63--135",
isbn="978-3-642-04458-8",
doi="10.1007/978-3-642-04458-8_2",
url="https://doi.org/10.1007/978-3-642-04458-8_2"
}

@article{caliman2025,
    title = {Improved detection of chaos with {L}agrangian descriptors using differential algebra},
    journal = {Physica D: Nonlinear Phenomena},
    volume = {472},
    pages = {134506},
    year = {2025},
    issn = {0167-2789},
    author = {Alexandru Căliman and Jérôme Daquin and Anne-Sophie Libert},
    doi = {https://doi.org/10.1016/j.physd.2024.134506}
}

@article{AntoniRuffo1995,
  author  = {Antoni, Mickael and Ruffo, Stefano},
  title   = {Clustering and Relaxation in {H}amiltonian Long-Range Dynamics},
  journal = {Physical Review E},
  volume  = {52},
  pages   = {2361--2374},
  year    = {1995},
  doi     = {10.1103/PhysRevE.52.2361}
}

@article{Chavanis2005,
  author  = {Chavanis, P. H. and Vatteville, J. and Bouchet, F.},
  title   = {Dynamics and thermodynamics of a simple model similar to self-gravitating systems: the HMF model},
  journal = {The European Physical Journal B - Condensed Matter and Complex Systems},
  year    = {2005},
  volume  = {46},
  number  = {1},
  pages   = {61--99},
  doi     = {10.1140/epjb/e2005-00234-0},
  url     = {https://doi.org/10.1140/epjb/e2005-00234-0},
  issn    = {1434-6036}
}

@article{CAMPA200957,
title = {Statistical mechanics and dynamics of solvable models with long-range interactions},
journal = {Physics Reports},
volume = {480},
number = {3},
pages = {57-159},
year = {2009},
issn = {0370-1573},
doi = {https://doi.org/10.1016/j.physrep.2009.07.001},
url = {https://www.sciencedirect.com/science/article/pii/S0370157309001586},
author = {Alessandro Campa and Thierry Dauxois and Stefano Ruffo}
}

@article{Pettini2024,
  author = {Pettini, Giulio and Gori, Matteo and Pettini, Marco},
  title = {From Geometry of {H}amiltonian Dynamics to Topology of Phase Transitions: A Review},
  journal = {Entropy},
  volume = {26},
  pages = {840},
  year = {2024},
  doi = {10.3390/e26100840}
}

@article{Manos02122011,
author = {Thanos Manos and Stefano Ruffo},
title = {Scaling with System Size of the Lyapunov Exponents for the {H}amiltonian Mean Field Model},
journal = {Transport Theory and Statistical Physics},
volume = {40},
number = {6-7},
pages = {360--381},
year = {2011},
publisher = {Taylor \& Francis},
doi = {10.1080/00411450.2011.651035},
URL = {https://doi.org/10.1080/00411450.2011.651035},
eprint = {https://doi.org/10.1080/00411450.2011.651035}
}

@article{Ginelli2011,
  author  = {Ginelli, Francesco and Takeuchi, Kazumasa A. and Chat{\'e}, Hugues and Politi, Antonio and Torcini, Alessandro},
  title   = {Chaos in the {Hamiltonian} Mean-Field Model},
  journal = {Physical Review E},
  volume  = {84},
  pages   = {066211},
  year    = {2011},
  doi     = {10.1103/PhysRevE.84.066211}
}

@book{Dauxois2002,
author="Dauxois, Thierry
and Latora, Vito
and Rapisarda, Andrea
and Ruffo, Stefano
and Torcini, Alessandro",
title="The {H}amiltonian Mean Field Model: From Dynamics to Statistical Mechanics and Back",
bookTitle="Dynamics and Thermodynamics of Systems with Long-Range Interactions",
year="2002",
publisher="Springer Berlin Heidelberg",
address="Berlin, Heidelberg",
pages="458--487",
isbn="978-3-540-45835-7",
doi="10.1007/3-540-45835-2_16",
url="https://doi.org/10.1007/3-540-45835-2_16"
}

@article{Dauxois2000,
  author  = {Dauxois, T. and Holdsworth, P. and Ruffo, S.},
  title   = {Violation of ensemble equivalence in the antiferromagnetic mean-field XY model},
  journal = {The European Physical Journal B - Condensed Matter and Complex Systems},
  year    = {2000},
  month   = aug,
  volume  = {16},
  number  = {4},
  pages   = {659--667},
  issn    = {1434-6036},
  doi     = {10.1007/s100510070183},
  url     = {https://doi.org/10.1007/s100510070183}
}

@misc{chavanis2010,
      title={Dynamical and thermodynamical stability of isothermal distributions in the HMF model}, 
      author={Pierre-Henri Chavanis},
      year={2010},
      eprint={1007.4916},
      archivePrefix={arXiv},
      primaryClass={cond-mat.stat-mech},
      url={https://arxiv.org/abs/1007.4916}, 
}

@article{Messer1982,
  author  = {Messer, Joachim and Spohn, Herbert},
  title   = {Statistical mechanics of the isothermal {Lane-Emden} equation},
  journal = {Journal of Statistical Physics},
  year    = {1982},
  volume  = {29},
  number  = {3},
  pages   = {561--578},
  issn    = {1572-9613},
  doi     = {10.1007/BF01342187},
  url     = {https://doi.org/10.1007/BF01342187}
}

@article{Barre2005,
  author  = {Barr{\'e}, Julien and Bouchet, Freddy and Dauxois, Thierry and Ruffo, Stefano},
  title   = {Large Deviation Techniques Applied to Systems with Long-Range Interactions},
  journal = {Journal of Statistical Physics},
  year    = {2005},
  month   = may,
  volume  = {119},
  number  = {3},
  pages   = {677--713},
  issn    = {1572-9613},
  doi     = {10.1007/s10955-005-3768-8},
  url     = {https://doi.org/10.1007/s10955-005-3768-8}
}

@article{BraunHepp1977,
  author  = {Braun, W. and Hepp, K.},
  title   = {The {V}lasov dynamics and its fluctuations in the $1/N$ limit of interacting classical particles},
  journal = {Commun. Math. Phys.},
  volume  = {56},
  number  = {2},
  pages   = {101--113},
  year    = {1977},
  doi     = {10.1007/BF01611497}
}

@article{Inagaki1993,
  author  = {Inagaki, Shogo},
  title   = {Thermodynamic Stability of Modified {K}onishi-{K}aneko System},
  journal = {Prog. Theor. Phys.},
  volume  = {90},
  number  = {3},
  pages   = {577--584},
  year    = {1993},
  doi     = {10.1143/ptp/90.3.577}
}

@article{Yamaguchi2004,
  author  = {Yamaguchi, Yoshiyuki Y. and Barr{\'e}, Julien and Bouchet, Freddy and Dauxois, Thierry and Ruffo, Stefano},
  title   = {Stability criteria of the {V}lasov equation and quasi-stationary states of the {HMF} model},
  journal = {Physica A},
  volume  = {337},
  number  = {1--2},
  pages   = {36--66},
  year    = {2004},
  doi     = {10.1016/j.physa.2004.01.041}
}

@article{Botet1982,
  author  = {Botet, R. and Jullien, R. and Pfeuty, P.},
  title   = {Size Scaling for Infinitely Coordinated Systems},
  journal = {Phys. Rev. Lett.},
  volume  = {49},
  number  = {7},
  pages   = {478--481},
  year    = {1982},
  doi     = {10.1103/PhysRevLett.49.478}
}

@article{Botet1983,
  author  = {Botet, R. and Jullien, R.},
  title   = {Large-size critical behavior of infinitely coordinated systems},
  journal = {Phys. Rev. B},
  volume  = {28},
  number  = {7},
  pages   = {3955--3967},
  year    = {1983},
  doi     = {10.1103/PhysRevB.28.3955}
}

@article{Touchette2009,
  author  = {Touchette, Hugo},
  title   = {The large deviation approach to statistical mechanics},
  journal = {Phys. Rep.},
  volume  = {478},
  number  = {1--3},
  pages   = {1--69},
  year    = {2009},
  doi     = {10.1016/j.physrep.2009.05.002}
}

@misc{NIST:DLMF,
  title        = {{NIST Digital Library of Mathematical Functions}},
  editor       = {Olver, F. W. J. and {Olde Daalhuis}, A. B.
                  and Lozier, D. W. and Schneider, B. I.
                  and Boisvert, R. F. and Clark, C. W.
                  and Miller, B. R. and Saunders, B. V.
                  and Cohl, H. S. and McClain, M. A.},
  publisher    = {National Institute of Standards and Technology},
  howpublished = {\url{https://dlmf.nist.gov/}},
  url          = {https://dlmf.nist.gov/10.29},
  year         = {2026},
  author       = {{National Institute of Standards and Technology}},
  note         = {Release 1.2.8, September 15, 2026,
                  Sec.~10.29: Recurrence Relations and Derivatives}
}

@article{Yoshida1990,
  author = {Yoshida, Haruo},
  title = {Construction of higher order symplectic integrators},
  journal = {Physics Letters A},
  volume = {150},
  number = {5--7},
  pages = {262--268},
  year = {1990},
  doi = {10.1016/0375-9601(90)90092-3}
}

@article{Wilson1971,
  author = {Wilson, Kenneth G.},
  title = {Renormalization Group and Critical Phenomena. {I}. {R}enormalization Group and the {Kadanoff} Scaling Picture},
  journal = {Physical Review B},
  volume = {4},
  pages = {3174--3183},
  year = {1971},
  doi = {10.1103/PhysRevB.4.3174},
  url = {https://journals.aps.org/prb/abstract/10.1103/PhysRevB.4.3174}
}

@article{Caiani1997,
  author = {Caiani, Lando and Casetti, Lapo and Clementi, Cecilia and Pettini, Marco},
  title = {Geometry of Dynamics, {Lyapunov} Exponents, and Phase Transitions},
  journal = {Physical Review Letters},
  volume = {79},
  pages = {4361--4364},
  year = {1997},
  doi = {10.1103/PhysRevLett.79.4361},
  eprint = {chao-dyn/9702011},
  archivePrefix = {arXiv},
  url = {https://arxiv.org/abs/chao-dyn/9702011}
}

@article{Caiani1998PRE,
  author = {Caiani, Lando and Casetti, Lapo and Clementi, Cecilia and Pettini, Giulio and Pettini, Marco and Gatto, Raoul},
  title = {Geometry of Dynamics and Phase Transitions in Classical Lattice {$\phi^4$} Theories},
  journal = {Physical Review E},
  volume = {57},
  pages = {3886--3899},
  year = {1998},
  doi = {10.1103/PhysRevE.57.3886}
}

@article{Casetti2000,
  author = {Casetti, Lapo and Pettini, Marco and Cohen, E. G. D.},
  title = {Geometric Approach to {H}amiltonian Dynamics and Statistical Mechanics},
  journal = {Physics Reports},
  volume = {337},
  pages = {237--341},
  year = {2000},
  doi = {10.1016/S0370-1573(00)00069-7}
}

@article{Rugh1997,
  author = {Rugh, Hans Henrik},
  title = {Dynamical Approach to Temperature},
  journal = {Physical Review Letters},
  volume = {78},
  pages = {772--774},
  year = {1997},
  doi = {10.1103/PhysRevLett.78.772},
  url = {https://journals.aps.org/prl/abstract/10.1103/PhysRevLett.78.772}
}

@article{Franzosi2012,
  author  = {Franzosi, Roberto},
  title   = {Geometric Microcanonical Thermodynamics for Systems with First Integrals},
  journal = {Physical Review E},
  volume  = {85},
  pages   = {050101},
  year    = {2012},
  doi     = {10.1103/PhysRevE.85.050101}
}

@book{Huang1987,
  author    = {Huang, Kerson},
  title     = {Statistical Mechanics},
  edition   = {2},
  publisher = {John Wiley \& Sons},
  address   = {New York},
  year      = {1987}
}

@article{Mancho2013,
  author = {Mancho, Ana M. and Wiggins, Stephen and Curbelo, Jezabel and Mendoza, Carolina},
  title = {Lagrangian Descriptors: A Method for Revealing Phase Space Structures of General Time Dependent Dynamical Systems},
  journal = {Communications in Nonlinear Science and Numerical Simulation},
  volume = {18},
  pages = {3530--3557},
  year = {2013},
  doi = {10.1016/j.cnsns.2013.05.002}
}

@article{WigginsGarciaGarrido2022,
  author = {Wiggins, Stephen and Garc{\'i}a-Garrido, V{\'i}ctor J.},
  title = {Painting the Phase Portrait of a Dynamical System with the Computational Tool of {L}agrangian Descriptors},
  journal = {Notices of the American Mathematical Society},
  volume = {69},
  number = {6},
  pages = {936--949},
  year = {2022},
  doi = {10.1090/noti2489}
}

@article{Lopesino2017,
  author = {Lopesino, Carlos and Balibrea-Iniesta, Francisco and Garc{\'i}a-Garrido, V{\'i}ctor J. and Wiggins, Stephen and Mancho, Ana M.},
  title = {A Theoretical Framework for {Lagrangian} Descriptors},
  journal = {International Journal of Bifurcation and Chaos},
  volume = {27},
  number = {01},
  pages = {1730001},
  year = {2017},
  doi = {10.1142/S0218127417300014},
  eprint = {1705.10213},
  archivePrefix = {arXiv},
  url = {https://arxiv.org/abs/1705.10213}
}

@article{Hillebrand2022,
  author = {Hillebrand, Malcolm and Zimper, Sebastian and Ngapasare, Arnold and Katsanikas, Matthaios and Wiggins, Stephen R. and Skokos, Charalampos},
  title = {Quantifying chaos using {Lagrangian} descriptors},
  journal = {Chaos},
  volume = {32},
  number = {12},
  pages = {123122},
  year = {2022},
  doi = {10.1063/5.0120889},
  eprint = {2208.09853},
  archivePrefix = {arXiv},
  url = {https://arxiv.org/abs/2208.09853}
}

@article{Zimper2023,
  author = {Zimper, Sebastian and Ngapasare, Arnold and Hillebrand, Malcolm and Katsanikas, Matthaios and Wiggins, Stephen R. and Skokos, Charalampos},
  title = {Performance of chaos diagnostics based on {Lagrangian} descriptors. Application to the {4D} standard map},
  journal = {Physica D: Nonlinear Phenomena},
  volume = {453},
  pages = {133833},
  year = {2023},
  doi = {10.1016/j.physd.2023.133833},
  eprint = {2305.00978},
  archivePrefix = {arXiv},
  url = {https://arxiv.org/abs/2305.00978}
}

@article{Daquin2022,
  author = {Daquin, J{\'e}r{\^o}me and P{\'e}denon-Orlanducci, R{\'e}mi and Agaoglou, Makrina and Garc{\'i}a-S{\'a}nchez, Guillermo and Mancho, Ana Maria},
  title = {Global dynamics visualisation from {Lagrangian} Descriptors. Applications to discrete and continuous systems},
  journal = {Physica D: Nonlinear Phenomena},
  volume = {442},
  pages = {133520},
  year = {2022},
  doi = {10.1016/j.physd.2022.133520},
  eprint = {2206.04467},
  archivePrefix = {arXiv},
  url = {https://arxiv.org/abs/2206.04467}
}

@article{JimenezLopezDelta2026,
  author = {Jim{\'e}nez-L{\'o}pez, Javier and Garc{\'i}a-Garrido, V{\'i}ctor J.},
  title = {A Simple Chaos Indicator Based on the {Lagrangian} Descriptor Difference of Neighboring Orbits},
  journal = {International Journal of Bifurcation and Chaos},
  volume = {36},
  pages = {2630018},
  year = {2026},
  doi = {10.1142/S0218127426300181},
  eprint = {2506.16660},
  archivePrefix = {arXiv},
  url = {https://arxiv.org/abs/2506.16660}
}

@misc{JimenezLopez2026,
  author = {Jim{\'e}nez-L{\'o}pez, Javier and Garc{\'i}a-Garrido, V. J.},
  title = {Trajectory arclength reveals chaos},
  year = {2026},
  eprint = {2602.07660},
  archivePrefix = {arXiv},
  primaryClass = {nlin.CD},
  doi = {10.48550/arXiv.2602.07660},
  url = {https://arxiv.org/abs/2602.07660}
}

@article{Firpo1998,
  author = {Firpo, Marie-Christine},
  title = {Analytic estimation of the {Lyapunov} exponent in a mean-field model undergoing a phase transition},
  journal = {Phys. Rev. E},
  volume = {57},
  pages = {6599--6603},
  year = {1998},
  doi = {10.1103/PhysRevE.57.6599},
  eprint = {chao-dyn/9803024},
  archivePrefix = {arXiv}
}

@article{Antoniazzi2007,
    author = {Antoniazzi, Andrea and Califano, Francesco
             and Fanelli, Duccio and Ruffo, Stefano},
    title = {Exploring the Thermodynamic Limit of {H}amiltonian Models: {C}onvergence to the {V}lasov Equation},
    journal = {Physical Review Letters},
    volume = {98},
    pages = {150602},
    year = {2007},
    doi = {10.1103/PhysRevLett.98.150602}
}

@article{Barre2006Vlasov,
    author = {Barr{\'e}, Julien and Bouchet, Freddy
             and Dauxois, Thierry and Ruffo, Stefano
             and Yamaguchi, Yoshiyuki Y.},
    title = {The {Vlasov} Equation and the
             {Hamiltonian} Mean-Field Model},
    journal = {Physica A: Statistical Mechanics
             and its Applications},
    volume = {365},
    number = {1},
    pages = {177--183},
    year = {2006},
    doi = {10.1016/j.physa.2006.01.005},
    eprint = {cond-mat/0511070},
    archivePrefix = {arXiv}
}

@article{Braun1977,
  author  = {Braun, W. and Hepp, K.},
  title   = {The {Vlasov} dynamics and its fluctuations in the $1/N$ limit of interacting classical particles},
  journal = {Communications in Mathematical Physics},
  year    = {1977},
  month   = jun,
  volume  = {56},
  number  = {2},
  pages   = {101--113},
  issn    = {1432-0916},
  doi     = {10.1007/BF01611497},
  url     = {https://doi.org/10.1007/BF01611497}
}

@book{Spohn1991,
  author    = {Spohn, Herbert},
  title     = {Large Scale Dynamics of Interacting Particles},
  series    = {Texts and Monographs in Physics},
  publisher = {Springer Berlin Heidelberg},
  address   = {Berlin, Heidelberg},
  year      = {1991},
  isbn      = {978-3-642-84371-6},
  doi       = {10.1007/978-3-642-84371-6},
  url       = {https://doi.org/10.1007/978-3-642-84371-6}
}

\end{document}